\documentclass{article}
\usepackage{spconf,amsmath,graphicx,hyperref}

\usepackage{amssymb}
\usepackage{booktabs}
\usepackage{multirow} 
\usepackage{soul} 
\usepackage{xcolor}

\title{Audio Contrastive-based Fine-tuning: \\Decoupling Representation Learning and Classification}
\name{
Yang Wang\textsuperscript{$\clubsuit$$\spadesuit$}, Qibin Liang\textsuperscript{$\heartsuit$}, Chenghao Xiao\textsuperscript{$\diamondsuit$}, Yizhi Li\textsuperscript{$\clubsuit$}, Noura Al Moubayed\textsuperscript{$\diamondsuit$}, Chenghua Lin\textsuperscript{$\clubsuit$}
}
\address{
\textsuperscript{$\clubsuit$}The University of Manchester, \textsuperscript{$\diamondsuit$}Durham University, \textsuperscript{$\heartsuit$}RFAI Technology, \textsuperscript{$\spadesuit$}Automated Analytics
}

\begin{document}
%
\maketitle
\begin{abstract}
Standard fine-tuning of pre-trained audio models couples representation learning with classifier training, which can obscure the true quality of the learned representations. In this work, we advocate for a disentangled two-stage framework that separates representation refinement from downstream evaluation. First, we employ a ``contrastive-tuning'' stage to explicitly improve the geometric structure of the model’s embedding space. Subsequently, we introduce a dual-probe evaluation protocol to assess the quality of these refined representations from a geometric perspective. This protocol uses a linear probe to measure global linear separability and a k-Nearest Neighbours probe to investigate the local structure of class clusters. 
Our experiments on a diverse set of audio classification tasks show that our framework provides a better foundation for classification, leading to improved accuracy. Our newly proposed dual-probing framework acts as a powerful analytical lens, demonstrating why contrastive learning is more effective by revealing a superior embedding space. 
It significantly outperforms vanilla fine-tuning, particularly on single-label datasets with a large number of classes, and also surpasses strong baselines on multi-label tasks using a Jaccard-weighted loss. Our findings demonstrate that decoupling representation refinement from classifier training is a broadly effective strategy for unlocking the full potential of pre-trained audio models. 
Our code will be publicly available. 
\end{abstract}

\section{Introduction}
\label{sec:intro}

The advancement of deep learning has revolutionised audio processing, with transfer learning from large pre-trained models becoming a de facto paradigm \cite{turian2022hear, la2024benchmarking, li2024mert}. The success of this paradigm hinges on adapting a general-purpose model to a downstream task, a process that should ideally produce a high-quality embedding space with desirable geometric properties. 
The conventional method for this adaptation is end-to-end fine-tuning, where a classification head is added to the model and the entire network is optimised with a cross-entropy loss. While effective, this approach provides only an indirect signal for structuring the embedding space \cite{la2024benchmarking}. The primary objective is to create a separable decision boundary for a specific classifier, not to explicitly shape the embeddings themselves. This can result in a space that is ``good enough'' for the task but may lack strong intra-class compactness and inter-class separation, potentially limiting robustness and transferability \cite{feng2024case}.

To address the limitations of indirect optimisation, we shift the focus to the direct geometric shaping of the embedding space. 
We investigate supervised contrastive learning not as a supplemental objective, but as the primary fine-tuning mechanism. 
Contrastive learning, in general, is explicitly geometric. For instance, the \textsc{SupCon} loss is designed to pull same-class embeddings together while pushing different-class embeddings apart, providing a direct signal to learn a well-structured space where semantic similarity corresponds to embedding proximity. 

Our primary contribution is two-fold. 
First, we propose \textsc{ConFit} (Contrastive Fine-tuning), a framework that fine-tunes pre-trained encoders using a supervised contrastive objective. To adapt this framework for the multi-label setting, we introduce the Jaccard-weighted Contrastive (\textsc{J-Con}) loss, a novel objective that handles partial label overlap. We demonstrate that this \textsc{ConFit} framework consistently outperforms vanilla end-to-end fine-tuning across diverse audio tasks. 
Second, and more critically, we introduce a comprehensive evaluation protocol designed to assess the intrinsic geometric quality of the resulting embedding space, moving beyond final accuracy scores. We argue that the quality of an encoder should be judged on its representations directly, not on the performance of a powerful, non-linear classifier that can act as a crutch for a poorly structured space. To this end, our protocol uses a dual-probe approach: a k-NN probe to assess local cluster compactness (evaluating the \textit{attraction} term of the loss) and a linear probe to assess global linear separability (evaluating the \textit{repulsion} term). This methodology provides a more rigorous measure of representation quality than standard end-to-end evaluation. 
Using this protocol, we demonstrate that for single-label classification tasks, \textsc{ConFit} constructs a geometrically superior embedding space compared to standard fine-tuning. The benefits are particularly pronounced on tasks with a large number of classes, a scenario where explicit geometric structuring is most critical.

\section{Methodology}
\label{sec:methodology}

A predominant paradigm in recent audio representation learning involves pre-training an encoder from scratch using either self-supervised or supervised objectives \cite{saeed2021contrastive, al2021clar}. 
In contrast to vanilla fine-tuning, we introduce \textsc{ConFit}, a framework that leverages the \textsc{SupCon} loss and its variants \cite{khosla2020supervised, feeney2023sincere} specifically as a transfer learning technique for encoders that have already been pre-trained. 
While a few prior studies \cite{hu2025label} have also applied contrastive learning to fine-tune pre-trained models, their evaluations have been limited to downstream task accuracy, offering limited insights into the resulting embedding space structure. To address this, our work additionally introduces a dual-probe framework which provides a direct geometric analysis. We investigate how \textsc{ConFit} reshapes the representation space, moving beyond simple performance scores to understand the structural improvements that lead to better performance. 

\subsection{Contrastive-tuning Stage}

The \textsc{ConFit} procedure begins by processing a batch of audio samples. For each audio sample $x_i$ in a given batch, we first pass it through a pre-trained audio encoder $f$. We use off-the-shelf encoders such as wav2vec2 \cite{baevski2020wav2vec}. This step yields a representation $z_i \in \mathbb{R}^{d}$ , where $d$ is the dimensionality of the embedding space. Following \cite{khosla2020supervised}, we then introduce a projector network $g$, that maps the representation $z_i$ into a new latent space where the contrastive loss is calculated. The final projected embedding is defined as $h_i$. This projector is used exclusively during the contrastive-tuning stage and is discarded once training is complete, ensuring that all downstream evaluations are performed using only the tuned encoder $f$.

The core of \textsc{ConFit} is driven by a supervised contrastive loss. To simplify the notation, we define the scaled cosine similarity between $h_i$ and $h_j$ as $s_{ij} = \text{sim}(h_i, h_j) / \tau$, where $\tau$ is a temperature hyper-parameter. The \textsc{SupCon} is defined for a batch of samples as:
\begin{equation}
    \sum_{i \in I} \frac{-1}{|P(i)|} \sum_{p \in P(i)} \log \frac{\exp(s_{ip})}{\sum_{k \in A(i)} \exp(s_{ik})}
\end{equation}
where $I$ is the set of all indices in the batch, $P(i)$ is the set of indices of positives, and $A(i) = I \setminus \{i\}$ is the set of all other indices. 
However, this formulation suffers from \textit{intra-class repulsion} \cite{feeney2023sincere}, as the denominator $\sum_{k \in A(i)}$ forces the anchor $h_i$ to be pushed away from \textit{all} other samples, including other positives. To address this, we also investigate the \textsc{Sincere} loss \cite{feeney2023sincere}, which modifies the denominator to only include true negatives:
\begin{equation}
    \sum_{i \in I} \frac{-1}{|P(i)|} \sum_{p \in P(i)} \log \frac{\exp(s_{ip})}{\exp(s_{ip}) + \sum_{n \in N(i)} \exp(s_{in})}
\end{equation}
where $N(i)$ is the set of negatives. This ensures that repulsion only occurs between samples of different classes. 

\textsc{SupCon} does not directly extend to the multi-label setting, where two samples may partially overlap in their labels. 
To address this, we propose the Jaccard-weighted Contrastive (\textsc{J-Con}) loss, which begins by replacing the binary notion of \textit{positiveness} with a Jaccard similarity that reflects the degree of label set overlap. Formally, for two samples $i$ and $j$ with label vectors $y_i$ and $y_j$:

\begin{equation}
    J(y_i, y_j) = \frac{|A_i \cap A_j|}{|A_i \cup A_j|} = \frac{y_i^T y_j}{\|y_i\|_1 + \|y_j\|_1 - y_i^T y_j}
\end{equation}
where $A_i$ is the set of active labels of sample $i$ and $\|y_i\|_1$ is the number of labels assigned to it. We define the \textsc{J-Con} loss by weighting each pair $(i,j)$ with its normalised Jaccard similarity and adjusting the denominator to only include the anchor and its true negatives:

\begin{equation}
    \sum_{i \neq j} \frac{-J(y_i, y_j)}{\sum_{p \neq i} J(y_i, y_p)} \log \frac{\exp(s_{ij})}{\exp(s_{ij}) + \sum_{k | J(y_i, y_k) \le \theta} \exp(s_{ik})}
\end{equation}

By defining and naming the \textsc{J-Con} loss, we establish it as a key component of our \textsc{ConFit} framework, allowing it to naturally handle multi-label supervision. 


\subsection{A Geometric Analysis with Dual-Probe Framework}

To complement the performance metrics from end-to-end fine-tuning, we introduce a dual-probe framework as an analysis lens. This provides deeper insight into the underlying geometric properties of the learned representations, allowing us to understand how the encoder structures the embedding space, not just the final downstream result. 
This framework is designed to diagnose the structure of the frozen embedding space after fine-tuning. It utilises two simple probes to assess geometric qualities. A \textbf{k-NN classifier} probes the \textit{local} structure of the space; high accuracy here reveals the formation of tight class clusters. In parallel, a \textbf{linear probe} assesses the \textit{global} structure by measuring the linear separability between these clusters. High linear probe accuracy indicates that the clusters are well-separated, which directly measures the efficacy of the contrastive repulsion term. By decoupling the evaluation into these two aspects, the dual-probe framework provides a more granular understanding of how our \textsc{ConFit} framework reshapes the representation space, connecting the mechanics of the loss function to the explicit geometric improvements that lead to better performance.

\section{Experiments}

\subsection{Datasets and Tasks}

\begin{table}[!t]\centering
\resizebox{\linewidth}{!}{ 
\begin{tabular}{lrrr}\toprule
\textbf{Dataset} &\textbf{Task} &\textbf{\# Classes} \\\midrule
EmoDB \cite{burkhardt2005database} &emotion classification &7 \\
MS-DB \cite{lostanlen2016deep} &instrument classification &8 \\
GTZAN \cite{tzanetakis2002musical} &music genre classification &10 \\
WMMS \cite{sayigh2016watkins} &mammal sound classification &31 \\
TIMIT \cite{garofolo1993timit} &speaker classification &630 \\
Libri \cite{panayotov2015librispeech} &speaker classification &2484 \\
\midrule
IRMAS \cite{bosch2018irmas} &instrument tagging &11 \\
FSD19 \cite{fonseca2019audio} &audio sound tagging &80 \\
\bottomrule
\end{tabular}
}
\caption{Statistics of datasets.}\label{tab:statistics}
\end{table}

We evaluate \textsc{ConFit} on a variety of tasks, including both speech and non-speech tasks. Table~\ref{tab:statistics} summarises the statistics of the eight audio datasets for our experiments.

\subsection{Implementation Details}

For all our experiments, we use the pre-trained wav2vec2 base model \cite{baevski2020wav2vec} as the audio encoder\footnote{While \textsc{ConFit} is model-agnostic, our focus is on the assessment for audio embeddings, not a comparative study of different encoders.}. All audio was segmented into 3-second chunks for training. The \textsc{ConFit} stage was run for 20 epochs with the AdamW optimiser and a cosine annealing learning rate schedule with linear warm-up. We performed a grid search for the batch size over $\{64, 128, 256, 512, 1024\}$ and for the peak learning rate over $\{10^{-5}, 5 \times 10^{-5}, 10^{-4}\}$. For the multi-label tasks, the Jaccard threshold $\theta$ was searched over $\{0.1, 0.2, 0.3, 0.4, 0.5\}$. The projector network $g$ was a 2-layer MLP (256-dim hidden, 128-dim output) with a ReLU activation, and the temperature $\tau$ for all contrastive losses was set to 0.07 \cite{khosla2020supervised}. To improve robustness, we applied online data augmentations with a probability of 0.8, including convolution with RIRs\footnote{\url{https://www.openslr.org/28}} and mixing with MUSAN\footnote{\url{https://www.openslr.org/17}} background noises. For evaluation, the encoder $f$ was frozen. A linear probe was trained for 40 epochs, while k-NN accuracy was computed using $k=5$ with cosine distance. We report accuracy for single-label tasks and mAP for multi-label tasks, with all scores averaged over 5 runs using different random seeds. For all our experiments, we use two NVIDIA A100 GPUs.

\section{Results and Discussion}

\subsection{Downstream Task Performance}

\begin{table*}[!t]
\centering
\begin{tabular}{lccccccccc}\toprule
\multirow{2}{*}{\textbf{Protocol}} &\multirow{2}{*}{\textbf{Method}} &\textbf{EmoDB} &\textbf{MS-DB} &\textbf{GTZAN} &\textbf{WMMS} &\textbf{TIMIT} &\textbf{Libri} &\multirow{2}{*}{\textbf{Avg}} \\
& & emotion & instrument & genre & sound & speaker & speaker & \\
\midrule
\textcolor{gray}{End-to-end} &\textcolor{gray}{Fine-tune} &\textcolor{gray}{82.25} &\textcolor{gray}{78.18} &\textcolor{gray}{72.41} &\textcolor{gray}{60.65} &\textcolor{gray}{67.62} &\textcolor{gray}{76.13} &\textcolor{gray}{72.87} \\
\midrule\midrule
\multirow{4}{*}{k-NN} & w/o Fine-tune &58.44 &59.15 &58.28 &67.74 &11.11 &30.30 &47.50 \\
& Fine-tune &\ul{77.49} &79.04 &72.07 &\ul{78.06} &73.73 &80.33 &76.79 \\
\cmidrule{2-9}
& \textsc{ConFit} (\textsc{SupCon}) &71.43 &\textbf{81.62} &\textbf{74.83} &\textbf{81.29} &\textbf{92.62} &\textbf{94.03} &\textbf{82.64} \\
& \textsc{ConFit} (\textsc{Sincere}) &\textbf{74.46} &\ul{78.88} &\ul{74.14} &\textbf{81.29} &\ul{91.35} &\ul{93.79} &\ul{82.32} \\
\midrule\midrule
\multirow{4}{*}{Linear Probe} &w/o Fine-tune &67.10 &64.72 &59.66 &34.84 &17.46 &57.58 &50.23 \\
&Fine-tune &\ul{80.52} &\textbf{77.26} &\textbf{72.07} &52.90 &80.32 &87.73 &75.13 \\
\cmidrule{2-9}
&\textsc{ConFit} (\textsc{SupCon}) &71.43 &\ul{76.65} &\ul{71.72} &\ul{56.13} &\ul{89.37} &\textbf{98.30} &\ul{77.27} \\
&\textsc{ConFit} (\textsc{Sincere}) &\textbf{81.39} &75.84 &70.69 &\textbf{60.65} &\textbf{91.35} &\ul{98.20} &\textbf{79.69} \\
\bottomrule
\end{tabular}
\caption{Main results. The best scores are \textbf{bold} and the second best ones are \ul{underlined}. 
}\label{tab:main}
\end{table*}

\begin{table}[!t]\centering
\resizebox{\linewidth}{!}{ 
\begin{tabular}{lrrrrr}\toprule
\textbf{Protocol} &\textbf{Method} &\textbf{IRMAS} &\textbf{FSD19} &\textbf{Avg} \\
\midrule
\multirow{3}{*}{k-NN} &w/o Fine-tune &21.30 &7.89 &14.60 \\
&Fine-tune &\ul{38.45} &\ul{26.58} &\ul{32.52} \\
\cmidrule{2-5}
&\textsc{ConFit} (\textsc{J-Con}) &\textbf{39.27} &\textbf{30.34} &\textbf{34.81} \\
\midrule
\multirow{3}{*}{Probe} &w/o Fine-tune &40.67 &25.64 &33.16 \\
&Fine-tune &\ul{55.44} &\ul{46.80} &\ul{51.12} \\
\cmidrule{2-5}
&\textsc{ConFit} (\textsc{J-Con}) &\textbf{57.56} &\textbf{48.31} &\textbf{52.94} \\
\bottomrule
\end{tabular}
}
\caption{Main results of representation quality evaluation on two audio multi-label classification datasets. 
}\label{tab:multilabel}
\end{table}

Our experimental results highlight the effectiveness of \textsc{ConFit}, with its benefits being most pronounced in specific, challenging scenarios.

\noindent\textbf{High-Cardinality Datasets Performance:} A key insight is that contrastive-tuning methods demonstrate an advantage on datasets with a large number of classes. As shown in Tables~\ref{tab:main}, this benefit is most significant for speaker classification on TIMIT (630 classes) and Libri (2484 classes). On these high-cardinality tasks, contrastive methods significantly outperform standard fine-tuning across both k-NN and linear probe evaluations. For instance, in the linear probe evaluation on Libri, \textsc{SupCon} achieves a score of 98.30, a nearly 11-point improvement over the 87.73 from standard fine-tuning. This suggests that the explicit repulsion term in the contrastive loss is effective at shaping a better embedding space with many distinct clusters, a scenario where the standard cross-entropy loss may struggle to enforce sufficient separation.

\noindent\textbf{Single-Label Task Performance:} On the broader set of single-label tasks, our dual-probe evaluation confirms the benefits of a decoupled approach. The k-NN evaluation (Table~\ref{tab:main} under k-NN protocol), which measures local cluster quality, shows that both \textsc{SupCon} and \textsc{Sincere} produce superior local embedding structures, outperforming the fine-tuning baseline by a significant margin on average. Similarly, the linear probe results (Table~\ref{tab:main}  under linear probe protocol), which assess global class separability, show that contrastive-tuned representations are more linearly separable, with \textsc{SupCon} and \textsc{Sincere} again achieving higher average scores than vanilla fine-tuning. These consistent improvements across different evaluation protocols validate our hypothesis that for single-label tasks, directly optimising the embedding geometry is a more effective strategy.

\noindent\textbf{Multi-Label Task Performance:} For the multi-label classification datasets (Table~\ref{tab:multilabel}), the results show that \textsc{J-Con} not only improves upon the baseline (without fine-tuning) but also consistently outperforms the vanilla fine-tuning method across both k-NN and linear probe evaluations. This indicates that \textsc{J-Con} successfully adapts the contrastive objective to the multi-label setting. 

\subsection{Dimensionality Contribution}

\begin{figure}[!t]
    \centering
    \includegraphics[width=1.0\linewidth]{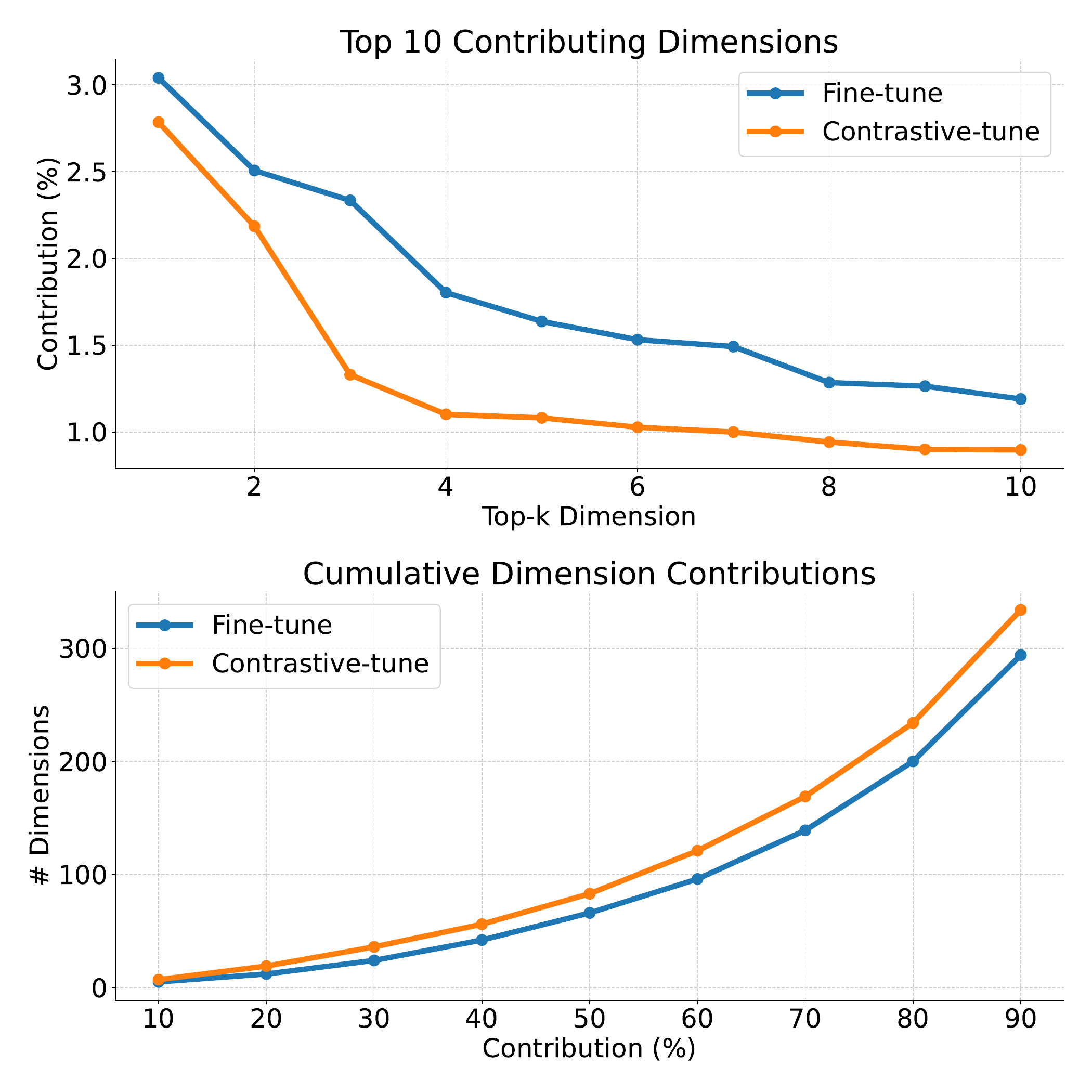}
    \caption{Dimensionality contribution of representations (fine-tuned vs contrastive-tuned). Upper: contribution percentages of the top 10 dimensions. Lower: number of dimensions required to reach 10\%–90\% of the cumulative contribution to the similarity metric.}
    \label{fig:rogue}
\end{figure}

Figure~\ref{fig:rogue} analyses the distribution of dimensional contributions, revealing how contrastive-tuning addresses the issue of \textit{rogue dimensions} \cite{timkey-van-schijndel-2021-bark, xiao-etal-2023-isotropy}. The upper plot shows that the k-th dimension in the standard fine-tuned model (blue line) holds a greater contribution percentage than its counterpart in the contrastive-tuned model (orange line), which shows a flatter, more even distribution of importance. 
The cumulative contribution plot (lower) reinforces this. To account for 50\% of the similarity metric, the fine-tuned model requires only 66 dimensions, whereas the contrastive model needs approximately 83. This demonstrates that contrastive objectives act as an effective regulariser. By preventing a small set of dimensions from dominating the representation, contrastive learning encourages a more isotropic distribution of semantic information. This property is desirable, as it ensures that the model is not reliant on a few features, which can improve the robustness of the learned representations for downstream tasks. This improved geometric structure likely contributes to the superior performance observed in the k-NN evaluations, where local neighbourhood structure is important.

\section{Conclusion}

In this work, we presented \textsc{ConFit}, a fine-tuning framework that decouples representation refinement from classifier training by using a dedicated supervised contrastive-tuning stage. We also introduced a dual-probe evaluation protocol to facilitate a more direct and geometrically-grounded assessment of audio representations. Our extensive empirical experiments demonstrate that this decoupled approach produces a superior embedding space across a diverse range of single- and multi-label classification tasks, leading to significant performance improvements over traditional end-to-end fine-tuning. 
We showed that contrastive-tuned representations exhibit both stronger local cluster compactness (via k-NN probe) and better global linear separability (via linear probe). Our analysis further revealed that these benefits are most pronounced on tasks with a high number of classes and that contrastive objectives promote a more isotropic embedding space. By incorporating a Jaccard-weighted loss, \textsc{ConFit} also outperforms fine-tuning baselines on multi-label tasks, confirming its broad applicability. These findings strongly advocate for a shift in focus during transfer learning: from solely optimising a final classifier to explicitly cultivating a well-structured representation space. 

\section{Disclosure of LLM Use}

We used the large language model (LLM) solely for polishing and proofreading the manuscript text. The LLM was not used to generate novel content, ideas, code, figures, or experimental results. All suggestions from the model were reviewed and verified by the authors to ensure accuracy and consistency. This use complies with the ICASSP policy on acceptable use of LLMs.

\vfill
\pagebreak

\bibliographystyle{IEEEbib}
\bibliography{strings,refs}

\end{document}